\documentclass[11pt,fleqn,twoside]{article}
\usepackage{amsfonts,latexsym}
\makeatletter
\newcommand{\prava}{\footnotesize\it
\begin{flushright}
\begin{minipage}{18cm}
Copyright \copyright 1998 by M. Ameduri and C.J. Efthimiou
\end{minipage}
\end{flushright}}

\newcommand{\name}[1]{\begin{flushleft}
                       \LARGE \bf #1
                       \end{flushleft}\vspace{-3mm}}

\newcommand{\Author}[1]{\begin{flushleft}
                       \it #1 \end{flushleft}}

\newcommand{\Adress}[1]{\begin{flushleft}
                       \it #1 \end{flushleft}}

\newcommand{\Date}[1]{\begin{flushleft}
                      \small  \it #1 \end{flushleft}}

\newcommand{\ehkol}{Author \ name}
\newcommand{\ohkol}{Article \ name}
\renewcommand{\@evenhead}{
\hspace*{-3pt}\raisebox{-15pt}[\headheight][0pt]{\vbox{\hbox to \textwidth
{\thepage \hfil \ehkol}\vskip4pt \hrule}}}
\renewcommand{\@oddhead}{
\hspace*{-3pt}\raisebox{-15pt}[\headheight][0pt]{\vbox{\hbox to \textwidth
{\ohkol \hfil \thepage}\vskip4pt\hrule}}}
\renewcommand{\@evenfoot}{}
\renewcommand{\@oddfoot}{}

\newcommand{\be}{\begin{equation}}
\newcommand{\ee}{\end{equation}}
\newcommand{\ba}{\hspace*{-5pt}\begin{array}}
\newcommand{\ea}{\end{array}}
\newcommand{\p}{\partial}
\newcommand{\ds}{\displaystyle}

\makeatother

\begin{document}
\setcounter{page}{132}

\thispagestyle{empty}

\renewcommand{\ehkol}{M. Ameduri and  C.J. Efthimiou}
\renewcommand{\ohkol}{Is the Classical Bukhvostov-Lipatov Model Integrable?}

\begin{flushleft}
\footnotesize \sf
Journal of Nonlinear Mathematical Physics \qquad 1998, V.5, N~2,
\pageref{ameduri-fp}--\pageref{ameduri-lp}. \hfill {\sc Letter}
\end{flushleft}

\vspace{-5mm}

{\renewcommand{\footnoterule}{}
{\renewcommand{\thefootnote}{}
 \footnote{\prava}}

\name{Is the Classical Bukhvostov-Lipatov\\
 Model Integrable?
A Painlev\'e Analysis}\label{ameduri-fp}

\Author{Marco AMEDURI~$\dag$ and  Costas J. EFTHIMIOU~$\ddag{}^*$}

\Adress{$\dag$~Newman Laboratory of Nuclear Studies, Cornell
University,\\
~~Ithaca, NY 14853,  USA \\
~~E-mail: marco@mail.lns.cornell.edu\\[2mm]
$\ddag$~Department of Physics and Astronomy,
        Tel Aviv University,\\
~~Tel Aviv, 69978, Israel\\[1mm]
${}^*$~Present address:\\
~~Department of Mathematics,  Harvard University, \\
~~Cambridge, MA 02138, USA\\
~~E-mail: costas@math.harvard.edu}

\Date{Received February 25, 1998; Accepted March 14, 1998}

\begin{abstract}
\noindent
In this work we apply the Weiss, Tabor and Carnevale integrability
criterion (Painlev\'e analysis) to the classical version of the
two dimensional Bukhvostov-Lipatov model. We are led to the conclusion
that the model is not integrable classically, except at a trivial
point where the theory can be described in terms of two uncoupled
sine-Gordon models.
\end{abstract}

\section{Introduction}
In a remarkable paper \cite{ameduri:BL}, Bukhvostov and Lipatov were able to
map the partition function for interacting instantons and anti-instantons
of the ${\rm O}(3)$ non-linear $\sigma$ model onto a two component
scalar f\/ield theory def\/ined by the Lagrangian
\begin{equation}
        {\cal L} = \sum_{i=1}^{2}
                \frac{1}{2} \partial_{\nu} \phi_{i}
                \partial^{\nu} \phi_{i} - \mu^{2}
                \cos (\lambda_{1} \phi_{1})
                \cos (\lambda_{2} \phi_{2}).
        \label{ameduri:Lag}
\end{equation}
They further showed that the quantum version of the model above is
exactly solvable provided the couplings $\lambda_{1}$, $\lambda_{2}$ are
constrained by the following relation:
\begin{equation}
        \frac{1}{\lambda_{1}^{2}} + \frac{1}{\lambda_{2}^{2}} =
        \frac{1}{\pi}.
        \label{ameduri:constraint}
\end{equation}
The integrability in this case was proved  via the bosonization
technique \cite{ameduri:boso} and the Bethe Ansatz \cite{ameduri:low}.

\pagebreak

We will call the model described by the Lagrangian (\ref{ameduri:Lag})
the Bukhvostov-Lipatov (BL) model\footnote[1]{Strictly speaking, the term is
usually used for the fermionic counterpart of Eq.(\ref{ameduri:Lag}) subject
to the constraint (\ref{ameduri:constraint}). However,
this slight abuse of terminology should not create any confusion.}.
The model (\ref{ameduri:Lag}) has been studied again recently in \cite{ameduri:saleur},
where other integrable cases have been found, which are fundamental to an
understanding of impurity problems in quantum wires. For a related model, see
also \cite{ameduri:fateev}.

An important open question, only partially addressed in \cite{ameduri:saleur},
concerns the integrability of the classical
equations of motion which can be derived from Eq.(\ref{ameduri:Lag}). This is
crucial to assess whether one can apply semiclassical considerations
to the corresponding quantum mechanical model. In the present
paper we analyze this problem via the Weiss, Tabor and Carnevale (WTC)
integrability criterion
\cite{ameduri:WTC}.

}

Some comments are due at this point. The Lagrangian
(\ref{ameduri:Lag}) can be regarded as being of the general form
(with complex couplings $\lambda_{a}$)
\begin{equation}
        {\cal L} = \sum_{a}
                \frac{1}{2} \partial_{\nu} \vec{\Phi}^{(a)} \cdot
                \partial^{\nu} \vec{\Phi}^{(a)} - \frac{\mu^2}{4}
                \prod_{a} \left\{
                \sum_{ \vec{\alpha}^{(a)}}\,
                e^{\lambda_{a}\, \vec{\alpha}^{(a)} \cdot \vec{\Phi}^{(a)}}
                \right\},
        \label{ameduri:generalL}
\end{equation}
where $\vec\alpha$ denotes the simple positive roots of the
Lie algebra $G$ and the index $a$ labels the dif\/ferent
algebras (possibly copies of the same one)
that appear in the potential. Obviously the well known
Toda Lagrangian is a special case of (\ref{ameduri:generalL}) for
$a=1$ and $G$ being a f\/inite or af\/f\/ine Lie algebra
\cite{ameduri:affine}.

A lot of work has been done in
the past twenty years in the f\/ield of integrable models. In particular,
the Toda f\/ield theories have been thoroughly studied, both in their
classical and quantum versions (for a review of the results see
\cite{ameduri:toda} and references therein).
In particular, Yoshida has shown \cite{ameduri:Y}
 that the integrable Toda f\/ield theories
are characterized by the Painlev\'e property.
More recently,
the authors of \cite{ameduri:GIM} used the WTC criterion to examine
integrability
of the hyperbolic Toda f\/ield theories for which other techniques have
not been applied so far. In \cite{ameduri:GIM}, the autors concluded that
the hyperbolic Toda f\/ield theories, although conformally invariant,
are {\it not} integrable since they fail the Painlev\'e test. The
BL model provides yet another example of theories belonging to the
general class def\/ined by (\ref{ameduri:generalL}).

The outline of the paper is the following:
f\/irst, in Section~\ref{ameduri:sec:2}, we brief\/ly review the WTC
algorithm and we apply it to the sine-Gordon (SG) model in a format
best suited for the problem to follow. In Section~\ref{ameduri:sec:3}, we
present its application
to the case of the BL model. Finally, Section~\ref{ameduri:sec:4}
contains a discussion of
the result.

\section{The Painlev\'e property for a PDE}
\label{ameduri:sec:2}

An ordinary dif\/ferential equation (ODE) is said to possess the Painlev\'e
property if all of its movable singularities are poles \cite{ameduri:booksODE}.
The connection between the Painlev\'e property and the integrability
of an ODE had been noted since the work of
S.~Kowalevskaya~\cite{ameduri:kowa} concerning the integrability of a
rotating rigid body.

The relation between integrability and the absence of movable critical
points was made more explicit through the work of Ablowitz, Ramani and
Segur~\cite{ameduri:ASR}, who established the following conjecture:
every ODE obtained by an exact reduction from a partial dif\/ferential
equation (PDE) solvable via the inverse scattering transform possesses
the Painlev\'e property. This led to the formulation of a
three-step algorithm
which allows one to test for the absence of multivalued
movable singularities in the solutions of a given ODE.

The def\/inition of the Painlev\'e property for PDEs and the
corresponding generalization of the aforementioned algorithm
was proposed by Weiss,
Tabor and Carnevale \cite{ameduri:WTC}.
This we will brief\/ly review in the following
subsection. For a comprehensive review see \cite{ameduri:RGB}.

\subsection{General description}

It is well known \cite{ameduri:osgo} that the singularities of a function
$f(z_{1}, z_{2}, \ldots, z_{n})$ of
$n>1$ complex variables cannot be isolated; rather they occur along
analytic manifolds of (complex) dimension $n-1$ determined by equations
of the form
\begin{equation}
        \chi \left( z_{1}, z_{2}, \ldots, z_{n} \right) = 0,
        \label{ameduri:singmani}
\end{equation}
$\chi$ being an analytic function of its variables in a neighborhood
of the singularity manifold def\/ined by Eq.(\ref{ameduri:singmani}).

One says \cite{ameduri:WTC} that a given PDE possesses the Painlev\'e
property if its solutions are single valued around the movable singularity
manifold~(\ref{ameduri:singmani}).

To test for the presence of the property one assumes that a solution
$u(z_{1}, z_{2}, \ldots, z_{n})$ of the PDE can be expanded around
the singularity manifold (\ref{ameduri:singmani}) as follows
\begin{equation}
        u = \chi^{-\alpha} \sum_{k=0}^{+\infty} u_{k} \chi^{k},
        \label{ameduri:defexp}
\end{equation}
where the coef\/f\/icients $u_{k}(z_{1}, z_{2}, \ldots, z_{n})$ are
analytic in a neighbourhood of $\chi=0$. One then substitutes the
above expansion~(\ref{ameduri:defexp}) in the PDE to determine the
value(s) of $\alpha$ and the recurrence relations\footnote[2]{The recurrence
relations are PDEs in the coef\/f\/icients $u_{k}$.} among the $u_{k}$'s.

If all the allowed values of $\alpha$ turn out to be integers
and the set of recurrence relations consistently allows for the
arbitrariness of initial conditions, then the given PDE is
said to possess the Painlev\'e property and is conjectured to
be integrable.

As an illustration of the method discussed above, we now analyze
the well known SG equation.

\subsection{An example: the sine-Gordon equation}

The two-dimensional SG equation \cite{ameduri:boso} arises as the
dynamical equation from the Lagrangian
\[
        {\cal L}_{{\rm SG}} = \frac{1}{2} \partial_{\nu} \phi
                \partial^{\nu} \phi - \mu^{2} \cos (\lambda \phi).
        \label{ameduri:sglag}
\]
In our notation $\nu=1,2$ and summation over repeated indices is
understood unless otherwise indicated. The metric is $g_{\mu \nu}
= {\rm diag}(1,-1)$. Also $\partial_{\nu} \equiv \partial/
\partial x^{\nu}$.
Introducing light-cone coordinates $x_{\pm} \equiv x_{1} \pm x_{2}$,
rescaling the f\/ield $\phi$ and f\/ixing the mass scale so that
$\mu^{2}=1$, we can write
\[
        \partial_{+} \partial_{-} \phi =
                \sin \phi.
        \label{ameduri:newSG}
\]

In order to apply the algorithm
we transform the equation above, following \cite{ameduri:GIM},
in the equivalent system\footnote[3]{The notation we are using to
discuss the simple SG case makes the problem more cumbersome
than usual \cite{ameduri:WTC}, but it establishes the conventions we
will adopt in the model we want to study.}:
\be
\label{ameduri:alleqsysSG}
 \dot{A} = BC, 
\qquad
 \dot{B} = - AC, 
\qquad C' = A,     
\ee
where the following new dependent variables have been def\/ined:
\[
        A \equiv \sin \phi, \qquad
        B \equiv \cos \phi,\qquad
        C \equiv \dot{\phi},
\]
and the following notation has been used:
\begin{equation}
        \dot{X} \equiv \partial_{-} X, \qquad
        X' \equiv \partial_{+} X.
        \label{ameduri:newnot}
\end{equation}

Substituting in the system of PDEs (\ref{ameduri:alleqsysSG}) a
series expansion for the functions $A$, $B$ and $C$ according to
(\ref{ameduri:defexp}) one f\/inds
\[
A = \chi^{-2} \sum\limits_{n=0}^{+\infty} A_{n} \chi^{n},
\quad
B = \chi^{-2} \sum\limits_{n=0}^{+\infty} B_{n} \chi^{n},
\quad
C = \chi^{-1} \sum\limits_{n=0}^{+\infty} C_{n} \chi^{n},
\]
with
\[
        A_{0} = 2i \dot{\chi} \chi', \qquad
        B_{0} = - 2 \dot{\chi} \chi', \qquad
        C_{0} = 2i \dot{\chi}.
\]
or
\[
        A_{0} = -2i \dot{\chi} \chi', \qquad
        B_{0} = - 2 \dot{\chi} \chi',
        C_{0} = -2i \dot{\chi}.
\]

We see that the solutions are single-valued around the singularity
manifold $\chi =0$. One f\/inds two possible singular behaviors for $A$,
corresponding to $e^{i \phi}$ or $e^{-i \phi}$ being singular.
The recurrence relations for the coef\/f\/icients are determined by
the following equations
\be
\label{ameduri:recuSG}
\ba{l}
\ds (n-2) \dot{\chi} A_{n} -
                C_{0} B_{n} - B_{0} C_{n} =
                \sum\limits_{m=1}^{n-1} B_{m} C_{n-m}  -
                \dot{A}_{n-1},
\\[3mm]
\ds (n-2) \dot{\chi} B_{n} + C_{0} A_{n}
                + A_{0} C_{n} =
                - \sum\limits_{m=1}^{n-1} A_{m} C_{n-m}  -
                \dot{B}_{n-1},
\\[3mm]
\ds (n-1) \chi' C_{n} - A_{n} = - C_{n-1}'.
\ea
\ee
One f\/inds that the determinant of the coef\/f\/icients vanishes
for $n=2$ and $n=4$. For these values one can check that
the relations (\ref{ameduri:recuSG})
vanish identically. This, together with the undeterminacy in $\chi$,
accounts for a complete set of boundary conditions.

We are now ready to discuss the central topic in the paper.

\section{The Bukhvostov-Lipatov model}
\label{ameduri:sec:3}

The dynamics of the model is characterized by the following
Lagrangian~\cite{ameduri:BL}:
\begin{equation}
        {\cal L}_{{\rm BL}} = \sum_{i=1}^{2}
                \frac{1}{2} \partial_{\nu} \phi_{i}
                \partial^{\nu} \phi_{i} - \mu^{2}
                \cos (\lambda_{1} \phi_{1})
                \cos (\lambda_{2} \phi_{2}).
        \label{ameduri:bllag}
\end{equation}
By appropriately f\/ixing
the mass scale, we can always set $\mu^{2} = 4$, so that the equations of
motion will read\footnote[4]{Notice that via an appropriate rescaling of
the f\/ields one can introduce a single coupling $\lambda_{1} / \lambda_{2}$.
For ease of notation we will keep writing the two couplings
separately.}
\be
\label{ameduri:BLeq}
\ba{l}
\partial_{+} \partial_{-} \phi_{1} =
                \lambda_{1} \sin (\lambda_{1} \phi_{1})
                \cos (\lambda_{2} \phi_{2}), \\[2mm]
\partial_{+} \partial_{-} \phi_{2} =
                \lambda_{2} \cos (\lambda_{1} \phi_{1})
                \sin (\lambda_{2} \phi_{2}).
\ea
\ee
Without loss of generality we can also assume that $\lambda_{1},
\lambda_{2}>0$.

As before, we introduce the new variables
\[
        A_{i} \equiv \sin (\lambda_{i} \phi_{i}) ,\qquad
        B_{i} \equiv \cos (\lambda_{i} \phi_{i}), \qquad
        C_{i} \equiv \partial_{-} \phi_{i}.
\]
where $i=1,2$ and the repeated index is not summed over.
With the notation def\/ined in Eq.(\ref{ameduri:newnot}) the equations of
motion will read
\be
\label{ameduri:BLsys}
\dot{A}_{i} = \lambda_{i} B_{i} C_{i}, \qquad
\dot{B}_{i} = - \lambda_{i} A_{i} C_{i}, \qquad
C_{1} ' = \lambda_{1} A_{1} B_{2}, \qquad  C_{2} ' = \lambda_{2} A_{2}
B_{1}.
\ee
We f\/ind the following allowed leading behaviors:
\[ \label{ameduri:Akleading}
A_{i} \sim A_{0}^{(i)} \chi^{-p_{i}},
\qquad B_{i} \sim B_{0}^{(i)} \chi^{-p_{i}}, \qquad C_{i} \sim
C_{0}^{(i)} \chi^{-1},
\]
where
\begin{equation}
        p_{1} \equiv \frac{2}{1 + \left(
        {\displaystyle \frac{\lambda_{2}}{\lambda_{1}}}
        \right)^{2}}, \qquad
        p_{2} \equiv \frac{2}{1 + \left(
        {\displaystyle \frac{\lambda_{1}}{\lambda_{2}}}
        \right)^{2}},
        \label{ameduri:pi}
\end{equation}
and the functions $B_{0}^{(1)}$, $B_{0}^{(2)}$ are constrained by the
relation
\begin{equation}
        B_{0}^{(1)} B_{0}^{(2)} = \frac{2\dot{\chi} \chi '}{
                \lambda_{1}^{2} + \lambda_{2}^{2} }.
        \label{ameduri:BBconstr}
\end{equation}

The leading coef\/f\/icients $A_{0}^{(i)}$, $B_{0}^{(i)}$, $C_{0}^{(i)}$
are additionally constrained. One can distinguish between four cases:

$\bullet$~{\bf Case I:}
\be
\ba{ll}
\ds  C_{0}^{(1)} = i \frac{p_{1}}{\lambda_{1}} \dot{\chi},
& \ds C_{0}^{(2)} = i \frac{p_{2}}{\lambda_{2}} \dot{\chi}, \\[3mm]
\ds  A_{0}^{(1)} = -i B_{0}^{(1)}, & \ds
                A_{0}^{(2)} = -i B_{0}^{(2)}.
\ea
                \label{ameduri:case1}
\ee

$\bullet$~{\bf Case II:}
        \be
\ba{ll}
\ds  C_{0}^{(1)} = i \frac{p_{1}}{\lambda_{1}} \dot{\chi},
& \ds C_{0}^{(2)} = -i \frac{p_{2}}{\lambda_{2}} \dot{\chi}, \\[3mm]
\ds   A_{0}^{(1)} = -i B_{0}^{(1)}, & \ds
 A_{0}^{(2)} = i B_{0}^{(2)}.
\ea
                \label{ameduri:case2}
\ee

$\bullet$~{\bf Case III:}
\be
\ba{ll}
\ds C_{0}^{(1)} = -i \frac{p_{1}}{\lambda_{1}} \dot{\chi}, &
\ds  C_{0}^{(2)} = i \frac{p_{2}}{\lambda_{2}} \dot{\chi}, \\[3mm]
\ds   A_{0}^{(1)} = i B_{0}^{(1)}, &
\ds    A_{0}^{(2)} = -i B_{0}^{(2)}.
\ea
                \label{ameduri:case3}
\ee

$\bullet$~{\bf Case IV:}
\be
\ba{ll}
\ds   C_{0}^{(1)} = -i \frac{p_{1}}{\lambda_{1}} \dot{\chi}, &
\ds C_{0}^{(2)} = -i \frac{p_{2}}{\lambda_{2}} \dot{\chi}, \\[3mm]
\ds  A_{0}^{(1)} = i B_{0}^{(1)}, &
\ds   A_{0}^{(2)} = i B_{0}^{(2)}.
\ea
                \label{ameduri:case4}
\ee

We notice f\/irst af all from Eq.(\ref{ameduri:pi})
that $p_{1}$ and $p_{2}$ are integer if and only if $\lambda_{1}
= \lambda_{2}$. In this case the equations (\ref{ameduri:BLeq}) can
be trivially
decoupled: the f\/ields $\phi_{\pm} \equiv \phi_{1} \pm \phi_{2}$
both obey the SG equation.

In general it is conjectured that for PDEs the original full Painlev\'e
property cannot be relaxed to its weaker form introduced in
\cite{ameduri:RDG} for two-dimensional dynamical systems, where the
leading behavior is allowed to be fractional.
It is important to notice though that a new choice of the dependent
variables might produce a pole type behavior: indeed, if $p_{1}= m/n$
(with $m<2n$), then $p_{2}=(2n-m)/n$; it could be that the
equations for the $n$-th
power of the functions $A_{i}$, $B_{i}$, $C_{i}$ satisfy the
full Painlev\'e property.

We therefore proceed to apply the second step of the algorithm.
We f\/irst notice that in all four cases
above [equations (\ref{ameduri:case1})--(\ref{ameduri:case4})] one of
the coef\/f\/icients $B_{k}^{(0)}$ is left undetermined. This allows
for a second initial condition arbitrariness, besides the position of
the singularity manifold.

Substituting in the equations (\ref{ameduri:BLsys}) an Ansatz of the type
(\ref{ameduri:defexp}), we f\/ind the following recursion relations for the
coef\/f\/icients (for $n > 0$):
\be
\label{ameduri:BLrecu}
\ba{l}
\ds \left(n - p_{i} \right) \dot{\chi} A_{n}^{(i)}
                - \lambda_{i}C_{0}^{(i)} B_{n}^{(1)}
                - \lambda_{i} B_{0}^{(i)} C_{n}^{(i)} =
                \lambda_{1} \sum \limits_{m=1}^{n-1}
                B_{n-m}^{(i)} C_{m}^{(i)} - \dot{A}_{n-1}^{(i)}
                \\[3mm]
\ds \left(n - p_{i} \right) \dot{\chi} B_{n}^{(i)}
                + \lambda_{i}C_{0}^{(i)} A_{n}^{(1)}
                + \lambda_{i} A_{0}^{(i)} C_{n}^{(i)} =
                - \lambda_{1} \sum \limits_{m=1}^{n-1}
                A_{n-m}^{(i)} C_{m}^{(i)} - \dot{B}_{n-1}^{(i)}
                \\[3mm]
\ds \left( n-1 \right) \chi ' C_{n}^{(1)}
                - \lambda_{1} B_{0}^{(2)} A_{n}^{(1)}
                - \lambda_{1} A_{0}^{(1)} B_{n}^{(2)} =
                \lambda_{1} \sum \limits_{m=1}^{n-1}
                A_{n-m}^{(1)} B_{m}^{(2)} - C_{n-1}^{(1)}
                \\[3mm]
\ds \left( n-1 \right) \chi ' C_{n}^{(2)}
                - \lambda_{2} B_{0}^{(1)} A_{n}^{(2)}
                - \lambda_{2} A_{0}^{(2)} B_{n}^{(1)} =
                \lambda_{2} \sum \limits_{m=1}^{n-1}
                A_{n-m}^{(2)} B_{m}^{(1)} - C_{n-1}^{(2)}
\ea
\ee
where the coef\/f\/icients $A_{0}^{(i)}$, $B_{0}^{(i)}$, $C_{0}^{(i)}$
are constrained by (\ref{ameduri:BBconstr})--(\ref{ameduri:case4}).

The set of recurrence relations (\ref{ameduri:BLrecu}) will
completely determine the coef\/f\/icients $A_{n}^{(i)}$,
$B_{n}^{(i)}$, $C_{n}^{(i)}$ unless
for some value of $n$ the determinant of the coef\/f\/icients
vanishes\footnote[5]{These values of $n$ are referred to as
resonances \cite{ameduri:RGB}.}.
In this latter case, for the Ansatz (\ref{ameduri:defexp})
to be correct the equations (\ref{ameduri:BLrecu}) must consistently
reduce to a set of trivial identities.

We therefore proceed to study the determinant $D$ of the coef\/f\/icients
associated with equation
(\ref{ameduri:BLrecu}). In all four cases
(\ref{ameduri:case1})--(\ref{ameduri:case4}) one
f\/inds
\[
        D = \left( \chi ' \right)^{2} \left( \dot{\chi} \right)^{4}
        (n+1)n(n-1)(n-2)(n-2p_{1})(n-2p_{2}).
\]
We therefore see that
for general values of $p_{1}$ and $p_{2}$ one f\/inds two resonances at
$n=1$ and $n=2$. One can check that the rank of the corresponding $6
\times 6$~matrices is f\/ive. This means that one is allowed for only
two extra arbitrary functions in the expansion (\ref{ameduri:defexp}).
This implies a total of only four arbitrary coef\/f\/icients, which is
insuf\/f\/icient to allow for complete arbitrariness of initial conditions.

We still have to examine the possibility that the values of the
coef\/f\/icients $\lambda_{1}$ and $\lambda_{2}$ are such that the
numbers $2p_{1}$, $2p_{2}$ lead to integer resonances. In this case
some appropriate power of $A_{i}$ and $B_{i}$ would have a pole-like
dominant behavior. This is the case provided
\[
        \left( \frac{\lambda_{1}}{\lambda_{2}} \right)^{2} =
        \frac{1}{3}, \quad 1, \quad  3.
\]
As mentioned before the case $(\lambda_{1}/\lambda_{2})^{2}=1$ is
trivially integrable (and consistently one f\/inds that the rank of
the coef\/f\/icient matrices at the resonances is small enough to accomodate
six initial conditions).

The other two cases lead to the appearence of an additional resonance
at $n=3$: if $(\lambda_{1}/\lambda_{2})^{2}=1/3$ or $3$, then the
determinant $D$ becomes
\[
        D= \left( \chi ' \right)^{2} \left( \dot{\chi} \right)^{4}
        (n+1)n(n-1)^{2}(n-2)(n-3).
\]
One f\/inds room for a new arbitrary coef\/f\/icient at the $n=3$ level.
The rank of the $n=1$ matrix though remains equal to one, so that
even for these values of the couplings the number of arbitrary
coef\/f\/icients is f\/ive, which is not enough to have integrability.

\section{Discussion}
\label{ameduri:sec:4}

The analysis above leads us to the conclusion that the BL
model does not pass the Painlev\'e test and is therefore
classically not integrable, except for the trivial case where
$\lambda_{1} = \pm \lambda_{2}$, where the model gives rise to two
uncoupled SG equations. This is to be contrasted to the quantum
theory of the same model which, as mentioned in the introduction, was
shown to be integrable for particular choices of the couplings.

This should not come too surprising, though.
Indeed, in the language of two-di\-men\-sio\-nal
conformal f\/ield theory
one can regard the Lagrangian (\ref{ameduri:bllag})
as a free f\/ield Lagrangian perturbed by a potential built out of a
combination of vertex operators. It is indeed well known
\cite{ameduri:zamo} that a two-dimensional conformal f\/ield theory
may remain integrable after some relevant
perturbation, which breaks the conformal invariance, is added.

\subsection*{Acknowledgements}
We would like to thank Prof. A.~LeClair for useful discussions;
M.A. also thanks him for advice and support.

\label{ameduri-lp}


\begin{thebibliography}{99}
\footnotesize

\bibitem{ameduri:BL} Bukhvostov A.P. and Lipatov L.N., {\it Nucl.
Phys.  B}, 1981, V.180, 116.

\bibitem{ameduri:boso} Coleman S., {\it Phys. Rev.  D}, 1975, V.11,
2088.\\
Mandelstam S., {\it Phys. Rev.  D}, 1975, V.11, 3026.

\bibitem{ameduri:low} Lowenstein H., in: Recent Advances
        in Field Theory and Statistical Mechanics, J.B.~Zuber and
        R.~Stora eds., Les Houches, 1984.\\
        Korepin V.E.,  Bogoliubov N.M. and  Izergin A.G.,
Quantum Inverse Scattering Method and Correlation
        Functions, Cambridge University Press, Cambridge, 1993.

\bibitem{ameduri:saleur} Lesage F., Saleur H. and  Simonetti P.,
 Tunneling in    Quantum Wires I: Exact Solution of the Spin Isotropic Case,
        Cond-mat/9703220.\\
        Lesage F., Saleur H. and  Simonetti P., Tunneling in
        Quantum Wires II: A New Line of IR Fixed Points,
        Cond-mat/9707131.

\bibitem{ameduri:fateev} Fateev V.A., {\it Nucl. Phys.  B}, 1996,
V.473[FS], 509.

\bibitem{ameduri:WTC}Weiss J., Tabor M. and Carnevale G., {\it J.
Math. Phys.}, 1983, V.24, 522.\\
        Weiss J., {\it J. Math. Phys.}, 1983, V.24, 1405.
\bibitem{ameduri:affine}
Kac V.G., Inf\/inite Dimensional Lie Algebras, Cambridge
        University Press, New York, 1990.\\
        Goddard P. and Olive D., {\it Int. J. Mod. Phys. A}, 1986, V.1,
        303.

\bibitem{ameduri:toda}Olshanetsky M.A. and  Perelomov A.M.,
        {\it Phys. Rep.}, 1981, V.71, 313. \\
        Olshanetsky M.A. and  Perelomov A.M.,
        {\it Phys. Rep.}, 1983, V.94, 313. \\
        Christe P. and Mussardo G., {\it Int. J. Mod. Phys. A},
1990, V.A,  4581.\\
Grisaru M.T., Lerda A., Penati S. and Zanon D., {\it Nucl. Phys. B},
1990, V.346, 264.\\
         Corrigan E., Lectures given at CRM-CAP Summer School on
        Particles and Fields '94, Banf\/f, Canada, 16-24 Aug., 1994.

\bibitem{ameduri:Y} Yoshida H., in: Non Linear Integrable Systems --
Classical  Theory and Quantum Theory, M.~Jimbo and T.~Miwa eds.,
         Proceedings of RIMS Colloquium 1981, World Scientif\/ic, 1983.

\bibitem{ameduri:GIM}Gebert R.W., Inami T. and Mizoguchi S., {\it
Int. J. Mod. Phys. A}, 1996, V.11, 5479.
\bibitem{ameduri:booksODE}Ince E.L.,  Ordinary Dif\/ferential Equations,
Dover, New York, 1956.\\
         Hille E., Ordinary Dif\/ferential Equations in the Complex
        Plane, Wiley, New York, 1976.

\bibitem{ameduri:kowa}  Kowalevskaya S., {\it Acta Math.}, 1889, V.12, 177.

\bibitem{ameduri:ASR} Ablowitz M.J., Ramani A. and  Segur H., {\it Lett. Nuovo
        Cimento}, 1978, V.23, 333.\\
Ablowitz M.J., Ramani A. and Segur H., {\it J. Math. Phys.},
1980, V.21, 715.\\
         Ablowitz M.J., Ramani A. and  Segur H., {\it J. Math. Phys.},
1980, V.21, 1006.

\bibitem{ameduri:RGB}Tabor M. and Gibbon J.D., {\it Physica D}, 1986,
V.18, 180.\\
Ramani A., Grammaticos B. and Bountis T., {\it Phys. Rep.}, 1989,
V.180, 159.

\bibitem{ameduri:osgo}Osgood W.F.,  Topics in the Theory of Functions of
        Several Complex Variables, Dover, New York, 1966.

\bibitem{ameduri:RDG}Ramani A., Dorizzi B. and Grammaticos B., {\it
Phys. Rev. Lett.}, 1982, V.49, 1539.

\bibitem{ameduri:zamo}Zamolodchikov A.B., {\it JETP Lett.}, 1987,
V.46, 160. \\
        Zamolodchikov A.B., in:  Advanced Studies in Pure
        Mathematics, V.19, M.~Jimbo, T.~Miwa and  A.~Tsuchiya eds., 1989.
\end{thebibliography}
\end{document}